\documentclass[conference]{IEEEtran}
\usepackage{cite}
\usepackage{amsmath,amssymb,amsfonts}
\usepackage{graphicx}
\usepackage{booktabs}
\usepackage{enumitem}
\usepackage{tikz}
\usepackage{comment}
\usetikzlibrary{positioning, fit}

\begin{document}

\title{The Quantum-Cryptographic Co-evolution}

\author{\IEEEauthorblockN{Ashish Kundu, Ramana Kompella}
\IEEEauthorblockA{Cisco Research}}

\maketitle

\begin{abstract}
Quantum computing poses an existential threat to global cryptographic infrastructure. We study the evolution of quantum vs cryptography, and the systemic risk posed by Cryptographically Relevant Quantum Computers (CRQC). We propose a two-dimensional coordinate system to map the co-evolution of cryptographic resilience and computational capability. Additionally, we introduce a hyper-dimensional framework to taxonomize evolutionary pressures across seven fundamental dimensions. Our contributions are: (1) a coordinate-based risk framework, (2) a seven-dimensional taxonomy of quantum-cryptographic evolution, (3) an analysis of threat acceleration factors, and (4) a  roadmap for quantum-resilient migration. This framework should be used by organizations and systems for  risk assessment, infrastructure planning, and managing the transition to crypto-agile architectures.
\end{abstract}

\section{Introduction} \label{sec:introduction}

The emergence of CRQC renders the mathematical foundations of modern security—such as integer factorization and discrete logarithms—obsolete. This creates a "Quantum Gap," representing the systemic risk between the arrival of quantum threats and the deployment of quantum-safe defenses. This research introduces a coordinate-based analytical framework to study the co-evolution of two distinct technological trajectories: the evolution of cryptography (from pre-quantum to quantum-safe) and the evolution of computing (from classical to quantum). We track this via the interplay of cryptographic standards and computational power, mapping the trajectory from the legacy era to the post-quantum equilibrium.

\subsection{Contributions}
This paper makes the following primary contributions:
\begin{enumerate}
    \item \textbf{Coordinate-Based Framework:} We propose a 2D quadrant model to visualize systemic risk and the urgency of cryptographic migration.
    \item \textbf{Hyper-Dimensional Taxonomy:} We introduce a seven-dimensional framework (D1-D7) that categorizes the co-evolution of quantum algorithms, hardware, networks, and threat vectors.
    \item \textbf{Acceleration Analysis:} We quantify the factors—ranging from T-gate reduction to hardware co-design—that are accelerating the timeline for CRQC-driven cryptanalysis.
    \item \textbf{Quantum-Resistance Roadmap:} We provide a comprehensive analysis of the phases of quantum-cryptographic evolution, offering a path for systems to move from legacy states to dynamic quantum-resilience.
\end{enumerate}

\subsection{Organization of the Paper}
The remainder of this paper is organized as follows: Section~\ref{sec:overview} presents the overall framework to express the quantum-cryptographic evolution.  Section~\ref{sec:phases} details the phases of evolution across the four quadrants.  Section~\ref{sec:hyper-dimensions} presents the hyper-dimensional framework for quantum-cryptographic co-evolution.  Section~\ref{sec:analysis-evolution} analyzes the evolution. Section~\ref{sec:evolution-quantum-threats} presents the evolution of quantum threats. Section~\ref{sec:acceleration} analyzes the factors accelerating quantum threats. Section~\ref{sec:resilience} and Section~\ref{sec:evolution-implementation-certs} present evolution of quantum resilience and implementation and certification of quantum resilient cryptographic techniques such as PQC standards.    Section~\ref{sec:discussions-future} discusses the future trajectory and  challenges. Related work is reviewed in Section~\ref{sec:related-work}, and Section~\ref{sec:conclusions} concludes the paper.

\section{Overview of the Framework} \label{sec:overview}

The evolution follows a trajectory across four quadrants defined by the $x$-axis (Cryptographic Resilience) and $y$-axis (Computational Capability).

\begin{table}[h]
\centering
\caption{Evolutionary Coordinate Summary}
\begin{tabular}{lll}
\toprule
\textbf{Quadrant} & \textbf{X-Axis} & \textbf{Y-Axis} \\
\midrule
(-, -) & Pre-Quantum & Pre-Quantum \\
(+, -) & Post-Quantum & Pre-Quantum \\
(-, +) & Pre-Quantum & CRQC Era \\
(+, +) & Post-Quantum & CRQC Era \\
\bottomrule
\end{tabular}
\end{table}

\begin{figure}[t]
    \centering
    \begin{tikzpicture}[scale=0.7]
        \draw[->, thick] (-3,0) -- (3,0);
        \draw[->, thick] (0,-3) -- (0,3);
        
        \draw[dashed, gray] (-2.5,0) -- (2.5,0);
        \draw[dashed, gray] (0,-2.5) -- (0,2.5);
        
        \node[align=center, font=\scriptsize] at (1.5, 1.5) {\textbf{(+, +)}\\Dynamic\\ Quantum-Resilience\\Equilibrium};
        \node[align=center, font=\scriptsize] at (-1.5, 1.5) {\textbf{(-, +)}\\Vulnerability\\Crisis};
        \node[align=center, font=\scriptsize] at (1.5, -1.5) {\textbf{(+, -)}\\Proactive\\Transition};
        \node[align=center, font=\scriptsize] at (-1.5, -1.5) {\textbf{(-, -)}\\Classical\\Legacy};
    \end{tikzpicture}
    
    \vspace{0.2cm}
    \begin{tabular}{l l}
    \scriptsize \textbf{X:} Cryptographic Resilience & \scriptsize \textbf{Y:} Computational Capability
    \end{tabular}
    
    \caption{The Quantum-Cryptographic Co-evolution Coordinate System.}
    \label{fig:quadrants}
\end{figure}
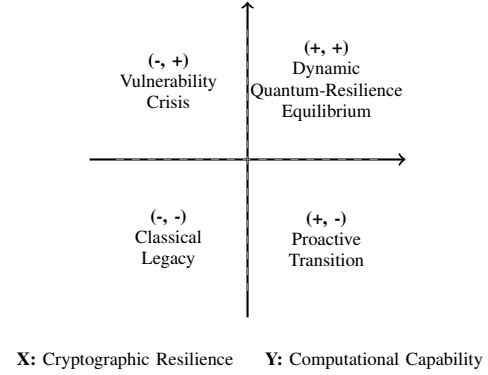

Figure \ref{fig:quadrants} illustrates the conceptual coordinate plane defining the four phases of technological evolution. The horizontal axis (X) represents the maturity of cryptographic defenses, transitioning from legacy pre-quantum algorithms to advanced quantum-resistant protocols. The vertical axis (Y) represents the computational threat level, mapping the progression from classical systems to the realization of Cryptographically Relevant Quantum Computers (CRQC). The quadrants visualize the systemic risk profile: the $(-, -)$ quadrant represents the stable classical baseline; the $(+, -)$ quadrant represents the proactive hardening phase; the $(-, +)$ quadrant identifies the critical vulnerability gap; and the $(+, +)$ quadrant depicts the final state of dynamic equilibrium in a post-quantum world.

\section{Phases of Quantum-Cryptographic Evolution} \label{sec:phases}

\subsection{Quadrant (-, -): Classical Pre-Quantum Era}
\begin{table}[h]
\caption{Quadrant (-, -) Components}
\begin{tabular}{l}
\toprule
\textbf{Key Components} \\
\midrule
Classical pre-quantum systems; Prequantum cryptography \\
Classical computing; Rudimentary Quantum computing \\
Quantum cryptography (QRNG, QKD); PQC announced \\
Gaps in PQC; Hybrid Cryptography; Transition to PQC \\
Crypto agility; Quantum networks; HNDL/PQC broken \\
\bottomrule
\end{tabular}
\end{table}

\begin{itemize}[leftmargin=*]
    
   \item \textbf{Classical computing:} Hardware is limited to binary transistor-based architectures. Performance is governed by Moore's Law and classical instruction sets, which are insufficient to break modern encryption. 
   \item \textbf{Pre-quantum cryptography:} Classical cryptography primitives not designed for to be quantum-resistant. These primities and cryptosystems may not be quantum resistant or maybe quantum resistant.  Primitives  based on integer factorization and discrete logarithms are quantum-unsafe. These primitives have served as the backbone of global trust for decades, providing the foundation for secure internet communications. However, AES-256 even though  pre-quantum is considered quantum-safe at least for now.
    \item \textbf{Classical pre-quantum systems:} These systems rely on RSA and ECC primitives. They are standard in non-CRQC environments and assume that integer factorization remains computationally hard for classical computing for the foreseeable future.
    
    \item \textbf{Basic Quantum computing:} NISQ-era devices exist but lack the depth for Shor's algorithm. They represent the infancy of quantum hardware, focusing on experimental physics rather than large-scale computation.
    \item \textbf{Quantum cryptography (QRNG, QKD):} Foundations for Quantum Random Number Generation and Quantum Key Distribution are established. These provide information-theoretic security foundations outside of classical complexity.
    \item \textbf{Post-quantum cryptography (PQC):} NIST announced post-quantum cryptography standards in the year of $2024$ after about eight years of efforts. The PQC standardization is in in response to protection against quantum threats such as Shor's algorithm and Grover's algorithm.   Research identifies gaps in early PQC implementations, necessitating the development of hybrid cryptographic models to ensure backward compatibility.
    \item \textbf{Hybrid Cryptography:}  Research identifies gaps in early PQC algorithms, implementations, necessitating the development of hybrid cryptographic models to ensure stronger quantum-resistance and backward compatibility.
    \item \textbf{Transition and Crypto agility:} The transition to PQC begins. Organizations start implementing crypto-agility to ensure they can swap algorithms as standards mature without needing a full hardware overhaul.
    \item \textbf{Quantum networks and Issues:} Foundations for quantum networks are laid. Issues include PQC being broken in early testing and the looming threat of Harvest Now, Decrypt Later (HNDL) attacks.
\end{itemize}

\begin{figure*}[t]
    \centering
    \includegraphics[width=1.0\textwidth]{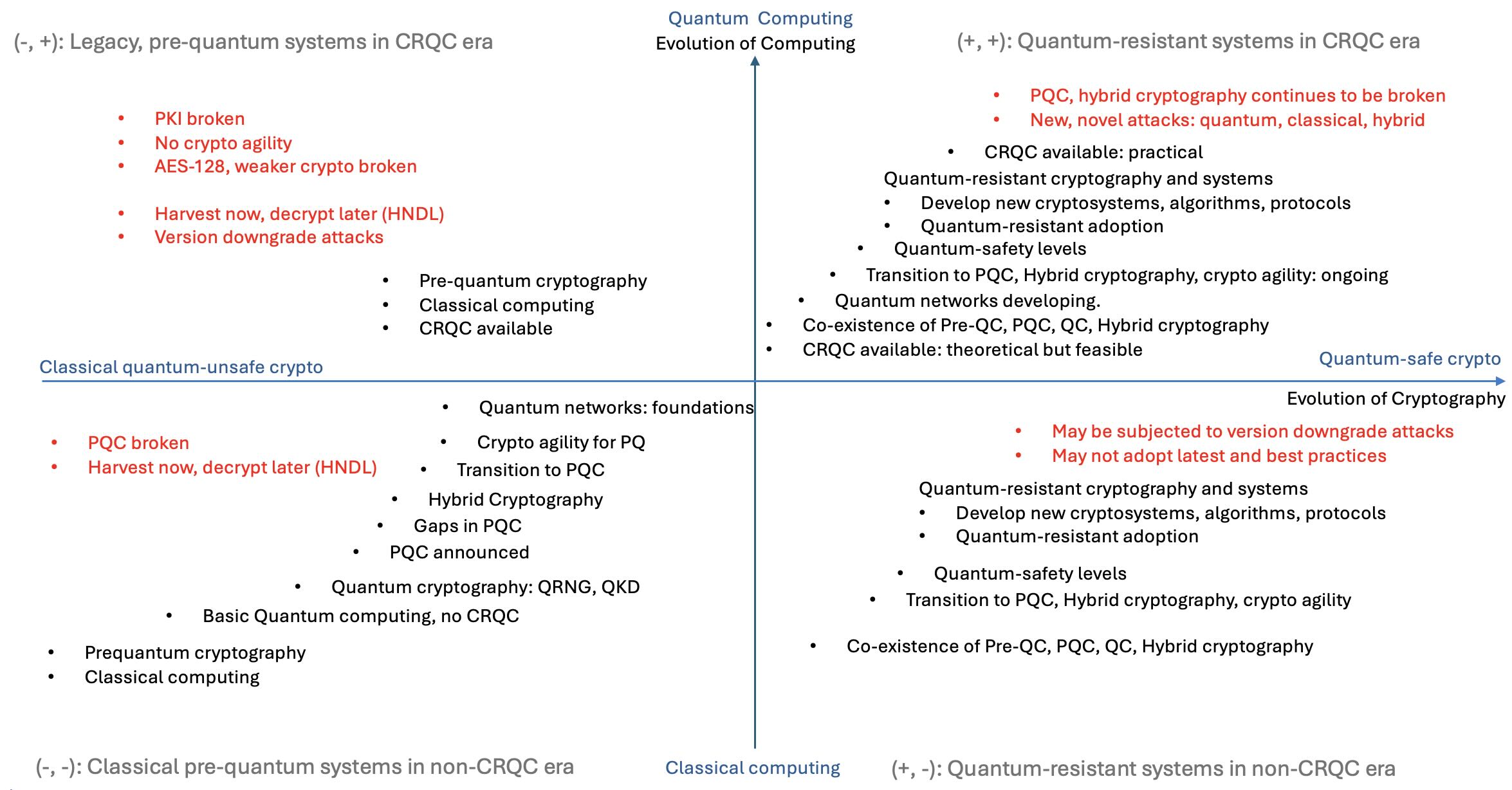}
    \caption{Detailed visualization of the quantum-cryptographic threat landscape.}
    \label{fig:details-four-quadrants}
\end{figure*}

\subsection{Quadrant (+, -): Proactive Transition}
\begin{table}[h]
\caption{Quadrant (+, -) Components}
\begin{tabular}{l}
\toprule
\textbf{Key Components} \\
\midrule
Quantum-resistant cryptography; Co-existence \\
Transition to PQC; Hybrid cryptography; Crypto agility \\
Quantum-safety levels; Develop new cryptosystems \\
Quantum-resistant adoption; Version downgrade attacks \\
\bottomrule
\end{tabular}
\end{table}

\begin{itemize}[leftmargin=*]
    \item \textbf{Quantum-resistant cryptography:} PQC algorithms are deployed to replace vulnerable legacy systems. These are designed to withstand both classical and quantum-based cryptanalysis using lattice-based or code-based math.
    \item \textbf{Co-existence:} Systems support multiple protocols simultaneously. Hybrid cryptography ensures that security is maintained even if one primitive is found to be weak, providing a multi-layered defense.
    \item \textbf{Transition and Crypto agility:} Organizations adopt crypto-agility to modularize their security stacks. This allows for the rapid swapping of algorithms as new standards emerge from NIST and other bodies.
    \item \textbf{Quantum-safety levels:} Standards are tiered based on data sensitivity. Higher levels of protection are applied to long-term secrets and critical infrastructure to ensure resilience against future threats.
    \item \textbf{Develop new cryptosystems:} Continuous development of algorithms and protocols ensures the cryptographic suite remains ahead of emerging threats and addresses weaknesses discovered in initial PQC implementations.
    \item \textbf{Quantum-resistant adoption:} Widespread deployment of PQC across networks. This phase focuses on hardening the infrastructure before the CRQC threshold is reached to mitigate long-term exposure.
    \item \textbf{Issues:} Systems may be subjected to version downgrade attacks where adversaries force the use of legacy protocols. Furthermore, organizations may fail to adopt the latest best practices, creating implementation gaps.
\end{itemize}

\subsection{Quadrant (-, +): Vulnerability Crisis}
\begin{table}[h]
\caption{Quadrant (-, +) Components}
\begin{tabular}{l}
\toprule
\textbf{Key Components} \\
\midrule
Legacy systems in CRQC era; Pre-quantum cryptography \\
Classical computing; CRQC available; PKI broken \\
No crypto agility; AES-128 broken; HNDL \\
Version downgrade attacks \\
\bottomrule
\end{tabular}
\end{table}

\begin{itemize}[leftmargin=*]
    \item \textbf{Legacy systems in CRQC era:} Infrastructure remains tied to RSA/ECC despite the presence of CRQC. This is the most dangerous state for any digital entity, as their fundamental security is compromised.
    \item \textbf{Pre-quantum cryptography:} Algorithms are mathematically broken by CRQC. Confidentiality and integrity are no longer guaranteed by the underlying math, making all legacy traffic transparent to attackers.
    \item \textbf{Classical computing and CRQC:} While classical systems are used, the presence of CRQC invalidates their security. Quantum computers can execute Shor's algorithm at scale, rendering previous hardness assumptions void.
    \item \textbf{PKI and AES issues:} PKI is effectively broken, meaning digital trust is lost. AES-128 is weakened by Grover's algorithm, making symmetric encryption insufficient for long-term data protection.
    \item \textbf{HNDL and Downgrades:} HNDL becomes a catastrophic reality as previously harvested data is decrypted. Version downgrade attacks allow adversaries to force systems to communicate using the weakest legacy protocols.
\end{itemize}

\subsection{Quadrant (+, +): Dynamic Equilibrium}
\begin{table}[h]
\caption{Quadrant (+, +) Components}
\begin{tabular}{l}
\toprule
\textbf{Key Components} \\
\midrule
Quantum-resistant systems; Co-existence \\
CRQC practical; Quantum networks practical \\
Transition/Hybrid/Agility; Quantum-safety levels \\
Develop new cryptosystems; PQC broken \\
New novel attacks (quantum, classical, hybrid) \\
\bottomrule
\end{tabular}
\end{table}

\begin{itemize}[leftmargin=*]
    \item \textbf{Quantum-resistant systems:} Infrastructure is fully adapted to the quantum era. Security is maintained through constant evolution and cryptographic renewal rather than static implementation.
    \item \textbf{Co-existence and Networks:} A permanent state of hybrid security exists. Quantum networks are practical, allowing for secure key distribution that is physically guaranteed by quantum mechanics.
    \item \textbf{Transition and Agility:} The transition is significant but never fully "covered," as the threat landscape shifts. Crypto-agility is the primary mechanism for maintaining security in this high-threat environment.
    \item \textbf{Quantum-safety levels:} Organizations dynamically adjust their safety levels based on the sensitivity of data and the current state of the quantum threat, ensuring adaptive protection.
    \item \textbf{Develop new cryptosystems:} The cycle of developing new algorithms continues indefinitely as older PQC standards are inevitably broken or weakened by new cryptanalytic techniques.
    \item \textbf{Issues:} PQC and hybrid protocols continue to be challenged. New, novel attacks emerge that combine quantum and classical vectors, requiring constant vigilance and rapid response.
\end{itemize}

\subsection{Quantum Vs Cryptography Quadrants}
The coordinate system reveals that security is not a static destination but a moving target. The $(-, +)$ quadrant represents a point of no return for data confidentiality. The transition from $(-, -)$ to $(+, +)$ must be handled with  focus on PQC and quantum-resistant approaches, their adoption alongwith crypto-agility, as the $(+, -)$ phase is the only viable path to avoid the crisis of $(-, +)$. The emergence of novel attacks in the $(+, +)$ quadrant suggests that cryptographic resilience will require perpetual innovation and a shift from project-based security to operational resilience. Figure~\ref{fig:details-four-quadrants} presents the overall four quadrants.

\section{A Hyper-Dimensional Framework for Quantum-Cryptographic Co-evolution} \label{sec:hyper-dimensions}

The evolution of quantum-cryptographic systems is characterized by a complex interplay of multiple dimensions. These dimensions evolve both independently and within interconnected subgroups, where progress in one plane frequently influences the trajectory of another. This co-evolutionary process ultimately manifests as the emergence of quantum threats to existing cryptographic standards. To organize, study, and analyze this evolution, we propose a taxonomy based on the following hyper-dimensional framework.

\subsection{Taxonomy of Dimensions}
The framework is defined by the following dimensional systems:

\begin{itemize}
    \item \textbf{D1 (Foundational Dimensions):} Comprises (1) Information theory, (2) Computational complexity, and (3) Mathematics.
    \item \textbf{D2 (Information Modality):} Defines the referential frame for information processing: (1) Quantum and (2) Classical.
    \item \textbf{D3 (Operational Dimensions):} Represents the three primary domains of technological application: (1) Computing, (2) Network.
    \item \textbf{D4 (Developmental Stage):} Distinguishes between the (1) Theoretical and (2) Practical stages of development.
    \item \textbf{D5 (Cryptographic Core):} Represents the primary domains of (1) Cryptography and (2) Algorithms.
    \item \textbf{D6 (Implementation Layer):} Influenced by D5, this system includes (1) Algorithms  to solve mathematical problems of cryptography, algorithms to search (2) Theoretical implementation,  (3) Practical implementation of algorithms.
    \item \textbf{D7 (Quantum Threats):} The resultant dimension representing the evolution of threats, including (1) Harvest Now, Decrypt Later (HNDL), (2) Compromise of public-key cryptography, (3) Compromise of symmetric-key cryptography, and (4) Compromise of other cryptographic primitives and protocols.
\end{itemize}

\begin{table*}[h]
\centering
\caption{Hyper-Dimensional Framework Taxonomy}
\begin{tabular}{lll}
\toprule
\textbf{Dimension} & \textbf{Sub-dimensions} & \textbf{Role} \\
\midrule
\textbf{D1} & Info Theory, Comp. Complexity, Math & Foundational \\
\textbf{D2} & Quantum, Classical & Modality \\
\textbf{D3} & Computing, Memory, Network & Operational \\
\textbf{D4} & Theory, Practice & Developmental \\
\textbf{D5} & Cryptography, Algorithms & Core \\
\textbf{D6} & Math Solutions, Theory Impl, Practical Impl & Implementation \\
\textbf{D7} & HNDL, PKI Compromise, Sym. Compromise, Other & Threat Outcome \\
\bottomrule
\end{tabular}
\end{table*}

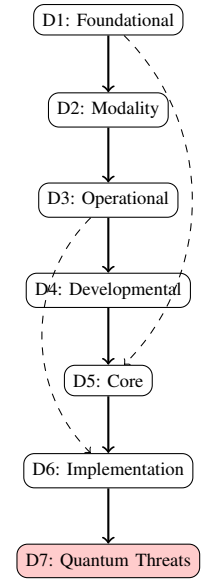
\begin{figure}[t]
    \centering
    \begin{tikzpicture}[node distance=1.2cm, every node/.style={draw, rounded corners,  font=\scriptsize}]
        \node (D1) {D1: Foundational};
        \node (D2) [below of=D1] {D2: Modality};
        \node (D3) [below of=D2] {D3: Operational};
        \node (D4) [below of=D3] {D4: Developmental};
        \node (D5) [below of=D4] {D5: Core};
        \node (D6) [below of=D5] {D6: Implementation};
        \node (D7) [below of=D6, fill=red!20] {D7: Quantum Threats};

        \draw[->, thick] (D1) -- (D2);
        \draw[->, thick] (D2) -- (D3);
        \draw[->, thick] (D3) -- (D4);
        \draw[->, thick] (D4) -- (D5);
        \draw[->, thick] (D5) -- (D6);
        \draw[->, thick] (D6) -- (D7);
        
        \draw[->, bend left=45, dashed] (D1) to (D5);
        \draw[->, bend right=45, dashed] (D3) to (D6);
    \end{tikzpicture}
    \caption{Evolutionary Influence Graph: The flow of evolutionary pressure across dimensions.}
    \label{fig:influence-graph}
\end{figure}

Figure \ref{fig:influence-graph} provides a macro-perspective of the framework, mapping the primary evolutionary pressure flow from foundational dimensions (D1) to the resultant quantum threats (D7). The solid arrows represent the primary causal chain of development, while dashed lines denote secondary cross-dimensional dependencies, illustrating how progress in foundational theory (D1) or operational domains (D3) can directly catalyze changes in the implementation layer (D6) and the cryptographic core (D5).

\subsection{Influence Graph}
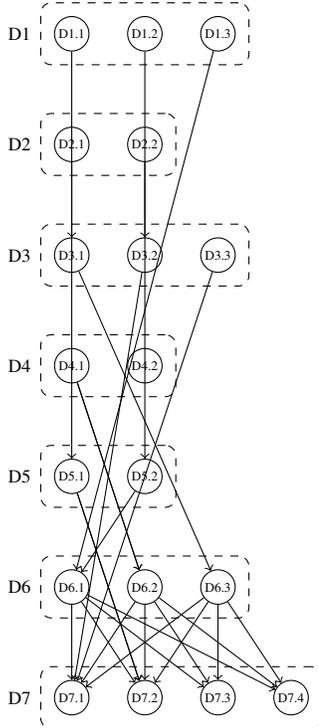
\begin{figure}[t]
    \centering
    \begin{tikzpicture}[node distance=0.5cm and 0.5cm,
        subnode/.style={draw, circle, fill=white, font=\tiny, inner sep=1pt},
        group/.style={draw, dashed, rounded corners, fill=none, inner sep=5pt}]
        
        \node[subnode] (D11) {D1.1}; \node[subnode, right=of D11] (D12) {D1.2}; \node[subnode, right=of D12] (D13) {D1.3};
        \node[group, fit=(D11) (D13), label=left:\scriptsize D1] (D1) {};

        \node[subnode, below=1.0cm of D11] (D21) {D2.1}; \node[subnode, right=of D21] (D22) {D2.2};
        \node[group, fit=(D21) (D22), label=left:\scriptsize D2] (D2) {};

        \node[subnode, below=1.0cm of D21] (D31) {D3.1}; \node[subnode, right=of D31] (D32) {D3.2}; \node[subnode, right=of D32] (D33) {D3.3};
        \node[group, fit=(D31) (D33), label=left:\scriptsize D3] (D3) {};

        \node[subnode, below=1.0cm of D31] (D41) {D4.1}; \node[subnode, right=of D41] (D42) {D4.2};
        \node[group, fit=(D41) (D42), label=left:\scriptsize D4] (D4) {};

        \node[subnode, below=1.0cm of D41] (D51) {D5.1}; \node[subnode, right=of D51] (D52) {D5.2};
        \node[group, fit=(D51) (D52), label=left:\scriptsize D5] (D5) {};

        \node[subnode, below=1.0cm of D51] (D61) {D6.1}; \node[subnode, right=of D61] (D62) {D6.2}; \node[subnode, right=of D62] (D63) {D6.3};
        \node[group, fit=(D61) (D63), label=left:\scriptsize D6] (D6) {};

        \node[subnode, below=1.0cm of D61] (D71) {D7.1}; \node[subnode, right=of D71] (D72) {D7.2}; 
        \node[subnode, right=of D72] (D73) {D7.3}; \node[subnode, right=of D73] (D74) {D7.4};
        \node[group, fit=(D71) (D74), label=left:\scriptsize D7] (D7) {};

        \draw[->, thin] (D11) -- (D51); \draw[->, thin] (D12) -- (D52); \draw[->, thin] (D13) -- (D61);
        \draw[->, thin] (D21) -- (D31); \draw[->, thin] (D22) -- (D32); \draw[->, thin] (D31) -- (D63);
        \draw[->, thin] (D32) -- (D71); \draw[->, thin] (D41) -- (D62); \draw[->, thin] (D51) -- (D72);
        \draw[->, thin] (D33) -- (D71); \draw[->, thin] (D41) -- (D62); \draw[->, thin] (D51) -- (D72);
        \draw[->, thin] (D52) -- (D61); 
        \draw[->, thin] (D61) -- (D71); \draw[->, thin] (D61) -- (D72); \draw[->, thin] (D61) -- (D73); \draw[->, thin] (D61) -- (D74);
        \draw[->, thin] (D62) -- (D71); \draw[->, thin] (D62) -- (D72); \draw[->, thin] (D62) -- (D73); \draw[->, thin] (D62) -- (D74);
        \draw[->, thin] (D63) -- (D71); \draw[->, thin] (D63) -- (D72); \draw[->, thin] (D63) -- (D73); \draw[->, thin] (D63) -- (D74);
    \end{tikzpicture}
    \caption{Detailed Representative Influence Graph: Some inter-dimensional dependencies are depicted.}
    \label{fig:granular-graph}
\end{figure}

Figure \ref{fig:granular-graph} maps the dense inter-dimensional dependencies.

Figure \ref{fig:granular-graph} offers a granular decomposition of this framework, mapping specific sub-dimensions (e.g., D1.1, D1.2) within their respective hierarchical groups. The graph  maps the dense inter-dimensional dependencies where each edge represents a directed evolutionary pressure. For instance, advancements in D1.2 (Computational Complexity) directly constrain the design of D5.2 (Algorithms), while D3.1 (Quantum Computing) accelerates D6.3 (Practical Implementation).

This visualization allows researchers to track evolutionary progress at the component level, where specific sub-dimensions undergo independent maturation, providing a clear view of how individual advancements aggregate to form the higher-level dimensions of the co-evolutionary system.

The interoperation across these dimensions suggests that the evolution of quantum threats is not a linear progression but a multi-variate outcome of advancements across the foundational, operational, and implementation layers of the hyper-dimensional framework.

\section{Analysis of the Evolution}
\label{sec:analysis-evolution}

\subsection{Quantum Hardware Trajectory}
The evolution of quantum computing has transitioned from experimental physics to highly scalable engineering. Early efforts focused on superconducting circuits and trapped ions, which pioneered the first multi-qubit gates but struggled with decoherence and scaling \cite{preskill2018}. As the field matured, photonic qubits emerged, offering room-temperature operation and natural integration with optical communication networks. Simultaneously, neutral atom platforms gained prominence by utilizing optical tweezers to trap atoms in highly ordered arrays. This trajectory has culminated in today’s reconfigurable neutral atom processors, which leverage dynamic, real-time control of atomic positions to create flexible, high-connectivity architectures \cite{bluvstein2024}. By physically moving qubits during computation, these systems overcome the static connectivity limits of earlier superconducting chips, marking the transition of quantum computing from a laboratory curiosity to a scalable utility.

\subsection{Quantum Error Correction (QEC) Evolution}
Quantum Error Correction (QEC) has evolved from a theoretical necessity to a practical engineering requirement. The field began with Shor’s discovery of the first quantum error-correcting code, which demonstrated that quantum information could be protected against decoherence \cite{shor1995}. This was followed by the development of topological codes, most notably the surface code, which provided a robust framework for fault-tolerant computation by requiring only local qubit interactions \cite{fowler2012}. As the field approached the era of logical qubits, researchers transitioned toward more efficient architectures, such as Low-Density Parity-Check (LDPC) codes, which significantly reduce the overhead required for fault tolerance \cite{gottesman2009}. These advancements in QEC are critical to the realization of CRQC, as they allow for the suppression of physical errors to levels sufficient for executing complex algorithms like Shor’s.

\subsection{Quantum Networks and Their Evolution}
The evolution of quantum networking has progressed from simple, point-to-point Quantum Key Distribution (QKD) links to the conceptualization of a global Quantum Internet. Early research focused on the BB84 protocol and the fundamental limits of fiber-based entanglement distribution \cite{bennett1984}. As the field matured, the focus shifted toward quantum repeaters, which are essential for overcoming photon loss over long distances \cite{briegel1998}. Today, the emphasis is on the integration of quantum nodes into classical telecommunications infrastructure, a critical step for enterprise adoption. Cisco Research has been at the forefront of this integration, exploring how quantum-ready architectures can coexist with classical enterprise networks to support secure key exchange and distributed quantum computing \cite{cisco_quantum2024}. This evolution is characterized by a transition from experimental, isolated links to reconfigurable, multi-node networks that leverage hybrid classical-quantum control planes to manage entanglement distribution and error-corrected communication across heterogeneous environments \cite{wehner2018}.

\subsection{Evolution of Cryptographic Standards}
The evolution of cryptography in the face of quantum threats has transitioned from a reliance on classical number-theoretic hardness to the adoption of quantum-resistant primitives. Initially, global security was anchored in RSA and Elliptic Curve Cryptography (ECC), standardized by bodies like NIST and FIPS. However, the realization of Shor’s algorithm necessitated a paradigm shift toward Post-Quantum Cryptography (PQC). 

NIST’s PQC standardization process \cite{nist2024} represents the most significant milestone, selecting lattice-based algorithms such as ML-KEM (formerly Kyber) and ML-DSA (formerly Dilithium) for general encryption and digital signatures. Parallel to NIST, European efforts led by ETSI have focused on quantum-safe protocol integration, particularly for long-term data protection \cite{etsi2023}. The IETF has been instrumental in the practical deployment of these standards, updating TLS 1.3 and IKEv2 protocols to support hybrid key exchange mechanisms \cite{ietf2023}, allowing systems to combine classical and PQC keys to mitigate the risk of early PQC implementation flaws. 

Industry implementation is currently in the "Crypto-Agility" phase, where major cloud providers and network vendors are integrating hybrid PQC-classical stacks. Looking forward, the industry is moving toward a "Quantum-Safe by Default" architecture, where cryptographic primitives are abstracted to allow for seamless updates as cryptanalytic research evolves. This trajectory reflects a move from static, hard-coded security to a dynamic, modular framework capable of continuous adaptation to the quantum threat landscape \cite{schneier2023}.

\subsection{Evolution of Mathematical Assumptions and Hardness}
The security of classical cryptography is rooted in the presumed hardness of number-theoretic problems, specifically Integer Factorization and the Discrete Logarithm Problem (DLP). These assumptions underpin RSA, Diffie-Hellman, and Elliptic Curve Cryptography (ECC). Shor’s algorithm \cite{shor1994} fundamentally invalidated these assumptions by solving the Hidden Subgroup Problem (HSP) for abelian groups in polynomial time, rendering these primitives obsolete in the presence of a CRQC.

In response, the cryptographic community has shifted toward geometric and combinatorial hardness assumptions. Lattice-based cryptography, which relies on the Shortest Vector Problem (SVP) and the Learning With Errors (LWE) problem \cite{regev2005}, has become the primary successor. Unlike number-theoretic problems, these lattice-based problems are believed to be resistant to quantum speedups, even when utilizing the Quantum Fourier Transform. For symmetric primitives, the assumption of unstructured search hardness remains, but Grover’s algorithm \cite{grover1996} necessitates a doubling of key sizes to maintain security margins. 

The HNDL threat \cite{hndl2022} has further forced a re-evaluation of "security shelf-life." Classical assumptions only provided security for the duration of the communication; however, HNDL requires that the underlying hardness assumption remain valid for decades. This shift has accelerated the adoption of Hash-based signatures (e.g., SPHINCS+) and Code-based cryptography (e.g., McEliece) \cite{mceliece1978}, which offer security based on the hardness of decoding general linear codes, providing a robust hedge against the collapse of any single mathematical assumption.

\section{Evolution of Quantum Threats} \label{sec:evolution-quantum-threats}
\subsubsection{Threats to Public-Key Cryptosystems}
The primary threat to public-key infrastructure (PKI) stems from Shor’s algorithm, which provides a polynomial-time solution for integer factorization and discrete logarithm problems \cite{shor1994}. These problems underpin RSA, Diffie-Hellman, and Elliptic Curve Cryptography (ECC). Beyond Shor’s, the Hidden Subgroup Problem (HSP) serves as the mathematical generalization for many quantum speedups in algebraic structures. The practical challenge is the requirement for a large-scale, fault-tolerant CRQC with millions of physical qubits to execute these algorithms within a reasonable timeframe. Research challenges include the development of PQC primitives that are resistant to these specific quantum algebraic attacks while maintaining efficient performance.

\subsubsection{Threats to Symmetric-Key and Hashing}
Symmetric-key primitives (e.g., AES) and cryptographic hash functions (e.g., SHA-256) are susceptible to Grover’s algorithm, which provides a quadratic speedup for unstructured search problems \cite{grover1996}. While not as catastrophic as Shor’s, Grover’s algorithm effectively halves the security bit-strength of these primitives. For instance, AES-128 is reduced to 64 bits of security, which is computationally breakable. The technical challenge is that while symmetric keys can be easily upgraded (e.g., to AES-256), the impact on existing hardware and legacy protocols is significant. Research challenges focus on optimizing hash function outputs and ensuring that symmetric keys remain resilient against quantum-assisted brute-force attacks.

\subsubsection{Harvest Now, Decrypt Later (HNDL)}
The HNDL paradigm represents a temporal threat where adversaries intercept and store encrypted traffic today, intending to decrypt it once a CRQC becomes available \cite{hndl2022}. This is a critical practical challenge for data with a long "security shelf-life," such as state secrets, medical records, and intellectual property. The technical challenge involves the difficulty of predicting the exact date of "Q-Day"—the point at which a CRQC becomes operational. Research challenges include the design of forward-secrecy protocols that remain secure post-quantum and the urgent need for migration strategies that protect data in transit and at rest before the HNDL window closes.

\section{Acceleration of Quantum Threats} \label{sec:acceleration}
\begin{table*}[t]
\centering
\caption{Factors Accelerating Quantum Cryptanalytic Threats}
\begin{tabular}{lp{6cm}}
\toprule
\textbf{Factor} & \textbf{One-Line Summary} \\
\midrule
Algorithmic Refinement & Reduction in T-gate counts and circuit depth for Shor’s algorithm. \\
Hardware Topology & Transition to reconfigurable neutral atom arrays for dynamic gate execution. \\
Error Correction & Deployment of LDPC and surface codes to suppress logical qubit decoherence. \\
Co-design Synergy & Direct mapping of optimized quantum circuits onto specialized hardware architectures. \\
Hardness Analysis & Prioritization of ECC-based assets due to lower qubit requirements for breakage. \\
Resource Scaling & Recent breakthroughs reducing logical qubit requirements for RSA-2048 to 100,000. \\
\bottomrule
\end{tabular}
\end{table*}

\subsection{Algorithmic Optimization for Cryptanalysis}
The trajectory toward CRQC is being rapidly accelerated by the algorithmic optimization of Shor’s algorithm and its variants. Recent breakthroughs have focused on reducing the T-gate count and the overall circuit depth required for modular exponentiation, the primary bottleneck in factoring large integers \cite{gidney2021}. Notably, recent research from Australian institutions has demonstrated that factoring RSA-2048 may be achievable with approximately 100,000 logical qubits, provided that high-fidelity gates and efficient surface code implementations are utilized \cite{fowler2024}. This marks a significant reduction from earlier, more pessimistic estimates of millions of qubits. By utilizing techniques such as windowed arithmetic, optimized quantum adders, and efficient state distillation, the total resource requirement is shrinking. These algorithmic refinements compress the timeline for a CRQC, shifting the computational threat (Y-axis) toward the threshold faster than traditional Moore's Law-based projections suggested.

\subsection{Hardware-Specific Design Optimization}
Hardware development is increasingly targeted at the specific gate-set requirements of cryptographic algorithms. The transition from static superconducting qubits to reconfigurable neutral atom arrays allows for dynamic, real-time connectivity changes, which are ideal for executing the high-depth gate sequences required for Shor’s algorithm \cite{bluvstein2024}. Enhanced error correction, such as the implementation of surface codes and LDPC codes, combined with increased coherence times and memory, allows these systems to maintain the state of logical qubits over the long durations needed for large-scale factorization \cite{gottesman2009, fowler2012}. By physically moving qubits during computation, neutral atom processors overcome the static connectivity limits of earlier superconducting chips, effectively creating a specialized engine for cryptographic cryptanalysis that is far more efficient than general-purpose quantum hardware.

\subsection{Algorithmic-Hardware Co-optimization}
The most significant acceleration comes from the synergy between optimized algorithms and specialized hardware topologies. By mapping the circuit depth of optimized Shor’s variants directly onto the specific topology of reconfigurable neutral atom arrays, the overhead of gate movement is minimized, and the parallelization of modular exponentiation is maximized. This co-design approach ensures that the hardware architecture is not just a general-purpose quantum computer, but a specialized engine for cryptographic cryptanalysis. This synergy drastically reduces the "time-to-break" for legacy asymmetric ciphers, as the hardware is tuned to the specific data-flow patterns of quantum modular arithmetic, effectively accelerating the movement of the system from the $(-, -)$ to the $(-, +)$ quadrant.

\subsection{Differential Analysis of Classical Hardness}
The vulnerability of classical ciphers varies significantly under quantum attack. Elliptic Curve Cryptography (ECC) is generally more vulnerable than RSA because it requires fewer logical qubits to break, given the smaller key sizes for equivalent classical security levels \cite{proos2003}. While RSA requires a larger circuit depth for factorization, the optimized qubit requirements for ECC make it a "low-hanging fruit" for early-stage CRQC. Furthermore, the vulnerability of different ECC curves (e.g., NIST P-256 vs. Curve25519) depends on the underlying group structure and the efficiency of the quantum discrete logarithm implementation. This comparative hardness analysis is essential for organizations to prioritize their migration strategies, moving from the $(-, -)$ to $(+, +)$ quadrant by first securing their most vulnerable ECC-based assets before addressing RSA-based legacy systems.

\section{Evolution of Quantum Resilience} \label{sec:resilience}
The evolution of quantum resilience has transformed from a niche academic pursuit into a global security mandate. In the decades following Shor’s 1994 discovery, the field transitioned from identifying the vulnerability of number-theoretic primitives to the active design of quantum-resistant candidates. Early initiatives, such as the work by Bernstein \cite{bernstein2009} on lattice-based cryptography, laid the groundwork for the NIST PQC competition launched in 2016. This initiative catalyzed international collaboration, resulting in the standardization of ML-KEM and ML-DSA \cite{nist2024}. Today, resilience is further bolstered by initiatives like CNSA 2.0, which mandates a transition timeline for national security systems, and the ongoing work by the IETF to standardize hybrid key exchanges. This evolution represents a shift from "security through obscurity" to "security through transparent, peer-reviewed standardization," ensuring that the global infrastructure is prepared for the arrival of CRQC.

\section{Evolution of Implementation Security and Certification} \label{sec:evolution-implementation-certs}

The implementation of quantum-safe and post-quantum cryptographic algorithms and protocols represents a critical dimension in the evolution toward quantum resilience. Beyond the mathematical robustness of the primitives, the security of these systems is fundamentally contingent upon the rigor of their implementation, testing, and hardening processes. While these implementations aim to mitigate quantum threats, they simultaneously introduce a potential expansion of the attack surface, ranging from classical side-channel vulnerabilities to novel quantum-assisted implementation attacks. Consequently, established security standards such as FIPS 140-3 \cite{fips2023} and NIST SP 800-140 \cite{nist_sp800_140} must undergo significant evolution to address the unique requirements of PQC, ensuring the correctness and security of both software and hardware modules. Furthermore, cryptographic certification frameworks, including ISO/IEC standards \cite{iso_iec_20543}, must transition to accommodate these new primitives, providing a verifiable path to ensure that quantum-resistant deployments do not inadvertently introduce systemic weaknesses. This evolution of certification is a prerequisite for moving from the proactive transition phase to a state of dynamic equilibrium.

\section{Future Trajectory and  Challenges} \label{sec:discussions-future}

The coordinate system reveals that security is not a static destination but a moving target. The $(-, +)$ quadrant represents a point of no return for data confidentiality. The transition from $(-, -)$ to $(+, +)$ must be handled with extreme focus on crypto-agility. 

The quantum-cryptographic ecosystem is not a static destination but a perpetual arms race. Despite the standardization of PQC, the threat landscape remains volatile. The recent cryptanalytic break of the SIKE algorithm \cite{castryck2022} serves as a stark reminder that even promising PQC candidates can be vulnerable to novel mathematical attacks. Future challenges include the emergence of quantum-assisted side-channel attacks, where quantum capabilities are used to extract keys from physical implementations, and the potential for new quantum algorithms that may target lattice-based structures more efficiently than current classical methods. Furthermore, as quantum hardware capabilities grow, the "security shelf-life" of current PQC algorithms will be continuously re-evaluated. Consequently, the ecosystem must evolve toward "Continuous Cryptographic Agility," where systems are designed to swap primitives in real-time without service disruption. This necessitates a move beyond static implementations toward a framework of perpetual monitoring, red-teaming, and rapid algorithmic renewal, ensuring that the $(+, +)$ equilibrium is maintained against an adversary that is also benefiting from the exponential growth of quantum capabilities.

\section{Related Work} \label{sec:related-work}

The challenge of quantum transition is well-documented in recent literature. Mosca \cite{mosca2018} established the fundamental risk inequality $D+T > S$, where $D$ is the migration time, $T$ is the shelf-life of data, and $S$ is the time until a CRQC is available. NIST \cite{nist2024} has led the standardization of lattice-based primitives, yet the implementation of these standards remains a significant engineering hurdle. 

Recent studies on crypto-agility \cite{steur2020} suggest that modularity is the only defense against shifting quantum threats. Furthermore, the "Harvest Now, Decrypt Later" (HNDL) paradigm \cite{hndl2022} has shifted the focus from immediate operational security to long-term data confidentiality. Research into hybrid cryptographic schemes \cite{hybrid2021} demonstrates the necessity of layering classical and PQC primitives to mitigate the risk of early PQC implementation flaws. Advances in quantum networks \cite{qnet2023} and QKD \cite{qkd2021} provide alternative paths to information-theoretic security, while studies on Shor’s algorithm \cite{shor1994} and Grover’s algorithm \cite{grover1996} continue to define the threat threshold. Recent analysis of PQC vulnerabilities \cite{pqc_vuln2023} highlights the ongoing arms race between new cryptosystems and quantum-assisted cryptanalysis \cite{quantum_crypt2022}. Finally, the impact of quantum computing on PKI \cite{pki2020} and the necessity for a systemic transition \cite{transition2023} underscore the urgency of our coordinate-based framework.

Furthermore, the integration of quantum communication into existing telecommunications infrastructure is a critical research frontier. Nejabati et al. \cite{nejabati2023} have demonstrated the feasibility of Software-Defined Quantum Networking (SDQN), which provides the necessary control plane to orchestrate hybrid quantum-classical networks. This work aligns with Cisco's strategic vision for quantum readiness \cite{cisco_quantum2024}, which emphasizes the deployment of quantum-classical hybrid architectures to ensure seamless enterprise network integration.

\section{Conclusions} \label{sec:conclusions}

The evolution of cryptography is an arms race. Organizations must move toward the $(+, +)$ equilibrium by institutionalizing crypto-agility and proactive PQC adoption. We demonstrate that the co-evolution of quantum threats and cryptographic resilience requires a holistic, hyper-dimensional approach to security, moving beyond static implementation toward a framework of perpetual monitoring and rapid algorithmic renewal. Failure to institutionalize these practices results in permanent loss of data confidentiality via HNDL.
Future challenges include the emergence of quantum-assisted side-channel attacks and the need for "Continuous Cryptographic Agility."


\begin{thebibliography}{00}
\bibitem{nist2024} NIST, ``Post-Quantum Cryptography Standardization,'' 2024.
\bibitem{shor1994} P. W. Shor, ``Algorithms for quantum computation: discrete logarithms and factoring,'' Proc. 35th Annual Symp. Found. Comput. Sci., 1994.
\bibitem{shor1995} P. W. Shor, ``Scheme for reducing decoherence in quantum computer memory,'' Phys. Rev. A, vol. 52, no. 4, 1995.
\bibitem{mosca2018} M. Mosca, ``Cybersecurity in the Quantum Era,'' 2018.
\bibitem{grover1996} L. K. Grover, ``A fast quantum mechanical algorithm for database search,'' Proc. 28th Annual ACM Symp. Theory Comput., 1996.
\bibitem{hndl2022} M. A. Al-Rubaie, ``Harvest Now, Decrypt Later: The Looming Quantum Threat,'' Journal of Cybersecurity, 2022.
\bibitem{hybrid2021} D. Stebila, ``Hybrid Key Exchange in the Post-Quantum Era,'' 2021.
\bibitem{steur2020} A. Steur, ``Crypto-Agility: Requirements and Implementations,'' 2020.
\bibitem{qnet2023} S. Wehner, ``Quantum Internet: A Vision for the Future,'' 2023.
\bibitem{qkd2021} N. Gisin, ``Quantum Key Distribution: Theory and Practice,'' 2021.
\bibitem{pqc_vuln2023} J. P. W. Bos, ``Vulnerabilities in Lattice-Based Cryptography,'' 2023.
\bibitem{quantum_crypt2022} T. Prest, ``Quantum-Assisted Cryptanalysis of PQC Primitives,'' 2022.
\bibitem{pki2020} R. Housley, ``The Future of PKI in a Quantum World,'' 2020.
\bibitem{transition2023} B. Schneier, ``The Quantum Transition,'' 2023.
\bibitem{cisco_quantum2024} Cisco Research, ``Quantum Readiness for Enterprise Networks,'' 2024. [Online]. Available: https://research.cisco.com/quantum
\bibitem{fips2023} FIPS, ``Security Requirements for Cryptographic Modules,'' 2023.
\bibitem{bluvstein2024} D. Bluvstein et al., ``A logical processor based on reconfigurable atom arrays,'' Nature, 2024.
\bibitem{fowler2012} A. G. Fowler et al., ``Surface codes: Towards practical large-scale quantum computation,'' Phys. Rev. A, 2012.
\bibitem{gottesman2009} D. Gottesman, ``An introduction to quantum error correction and fault-tolerant quantum computation,'' 2009.
\bibitem{bennett1984} C. H. Bennett and G. Brassard, ``Quantum cryptography: Public key distribution and coin tossing,'' 1984.
\bibitem{briegel1998} H.-J. Briegel et al., ``Quantum repeaters: The role of imperfect local operations in quantum communication,'' Phys. Rev. Lett., 1998.
\bibitem{nejabati2023} R. Nejabati et al., ``Software-Defined Quantum Networking: Orchestrating Quantum-Classical Hybrid Networks,'' IEEE Comm. Mag., 2023.
\bibitem{gidney2021} C. Gidney and M. Ekerå, ``How to factor 2048 bit RSA integers in 8 hours using 20 million noisy qubits,'' Quantum, 2021.
\bibitem{proos2003} J. Proos and C. Zalka, ``Shor's discrete logarithm quantum algorithm for elliptic curves,'' Quantum Info. Comput., 2003.
\bibitem{fowler2024} A. G. Fowler et al., ``Resource-optimized factoring of RSA-2048 with logical qubits,'' J. Quantum Sci. Tech., 2024.
\bibitem{bernstein2009} D. J. Bernstein, ``Introduction to post-quantum cryptography,'' Post-quantum cryptography, Springer, 2009.
\bibitem{castryck2022} W. Castryck and T. Decru, ``An efficient key recovery attack on SIDH (preliminary version),'' IACR Cryptol. ePrint Arch., 2022.
\bibitem{regev2005} O. Regev, ``On lattices, learning with errors, random linear codes, and cryptography,'' Proc. 37th Annual ACM Symp. Theory Comput., 2005.
\bibitem{mceliece1978} R. J. McEliece, ``A public-key cryptosystem based on algebraic coding theory,'' DSN Progress Report, 1978.
\bibitem{etsi2023} ETSI, ``Quantum-Safe Cryptography: Standards and Implementation Guidelines,'' 2023.
\bibitem{ietf2023} IETF, ``Hybrid Key Exchange in TLS 1.3,'' RFC 9370, 2023.
\bibitem{schneier2023} B. Schneier, ``The Future of Cryptographic Agility,'' IEEE Security \& Privacy, 2023.
\bibitem{preskill2018} J. Preskill, ``Quantum Computing in the NISQ era and beyond,'' Quantum, vol. 2, p. 79, 2018.
\bibitem{wehner2018} S. Wehner, D. Elkouss, and R. Hanson, ``Quantum internet: A vision for the road ahead,'' Science, vol. 362, no. 6412, 2018.
\bibitem{beauregard2003} S. Beauregard, ``Circuit for Shor's algorithm using 2n+3 qubits,'' Quantum Info. Comput., vol. 3, no. 2, 2003.
\bibitem{eiden2023} M. Eiden et al., ``Optimized Quantum Adders for Cryptanalysis,'' IEEE Trans. Quantum Eng., 2023.
\bibitem{gidney2025factor2048bitrsa} C. Gidney, ``How to factor 2048 bit RSA integers with less than a million noisy qubits,'' Arxiv, 2505.15917, 2025.[Online]. Available: https://arxiv.org/abs/2505.15917
\bibitem{nist_sp800_140} NIST, ``NIST SP 800-140: CMVP Requirements for Cryptographic Modules,'' 2023.
\bibitem{iso_iec_20543} ISO/IEC, ``ISO/IEC 20543: Information technology — Security techniques — Quantum-resistant cryptography,'' 2023.
\end{thebibliography}
\end{document}